\documentstyle[12pt]{article}
\input{psfig}
\long\def\comment#1{}
\begin{document}
\title{Y\={a}j\~{n}avalkya and the Origins of Pur\={a}\d{n}ic Cosmology}

\author{Subhash Kak\\
Department of Electrical \& Computer Engineering\\
Louisiana State University\\
Baton Rouge, LA 70803-5901, USA\\
Email: {\tt kak@ee.lsu.edu}}

\maketitle

\begin{abstract}
This paper shows that 
characteristic features
of Pur\={a}\d{n}ic cosmology, such as 
alternating cosmic ``continents'' and ``oceans'' of
successively doubling areas, can be traced to Vedic texts.
The \d{R}gveda speaks of seven regions of the universe,
and
Y\={a}j\~{n}avalkya, in B\d{r}had\={a}ra\d{n}yaka Upani\d{s}ad,
presents a cosmology that has all the essential features of the
Pur\={a}\d{n}ic system. This discovery solves the old puzzle of
the origin of 
Pur\={a}\d{n}ic astronomy. \\

\end{abstract}

\section*{Introduction}

The prehistory of Pur\={a}\d{n}ic astronomy is not
well understood. Although it is known that the Pur\={a}\d{n}as
contain very old material, some modern historians
of astronomy have
believed that the cosmology
presented there has no Vedic antecedents. In this
paper, we show that this belief is wrong, and 
a system similar to Pur\={a}\d{n}ic cosmology
is described in the Vedic tradition.

In the past few years a new understanding of the origins
of Indian astronomy has emerged.
In various publications,$^1$
we have sketched a history of Indian astronomy
from the earliest Vedic conceptions--- as expressed
in the astronomy of geometric altars--- to the
classical Siddh\={a}ntic astronomy. Meanwhile, the use
of modern computer packages has made it possible
to reexamine the astronomical references in the early
texts.$^2$
Although
recent work showed that Pur\={a}\d{n}ic and
Siddh\={a}ntic astronomies have several
common points, the origins of Pur\={a}\d{n}ic astronomy remained
unclear.

Here we present 
Y\={a}j\~{n}avalkya's
cosmology from 
B\d{r}had\={a}ra\d{n}yaka Upani\d{s}ad (BU) 3.3.2
which has hitherto escaped scholarly attention.
We show that this cosmology has all the basic
features of the later
Pur\={a}\d{n}ic cosmology and so it may be
viewed as the original source.

The great Upani\d{s}adic sage
Y\={a}j\~{n}avalkya
is a major figure in the earliest Indian
astronomy, that precedes Lagadha's
Ved\={a}\.{n}ga Jyoti\d{s}a.
Elsewhere, we have noted that in the
\'{S}atapatha Br\={a}hma\d{n}a, Y\={a}j\~{n}avalkya
speaks of a 95-year intercalary cycle to harmonize
the lunar and solar years,$^3$ and he describes the
non-uniform motion of the Sun.$^4$ We showed that
these were significant links in the early development
of Vedic astronomy. 
The 19th century European Indologists tended
to assign him to around 800 B.C., but a
reexamination of the evidence suggests
that he should be assigned to around 1800 B.C.
However, the chronology of Y\={a}j\~{n}valkya is not our concern here,
so will not comment any further on it; this question is discussed
at length elsewhere.$^5$

\section*{Vedic and Pur\={a}\d{n}ic Cosmological Ideas}

In brief,
the Vedas take the universe to be infinite in size. 
The universe is visualized in the image of the
cosmic egg, {\it Brahm\={a}\d{n}\d{d}a}.
Beyond our own universe lie other universes. 

The Pa\~{n}cavi\d{m}\'{s}a
Br\={a}hma\d{n}a 16.8.6 states that the heavens are
1000 earth diameters away
from the Earth.
The Sun was taken to be halfway to the heavens, so
this suggests a distance to the Sun to be about
500 earth diameters from the Earth.

Yajurveda, in the mystic hymn 17, dealing with the nature of the universe,
counts numbers in powers of
ten upto $10^{12}$. It has been suggested that this 
is an estimate of the size of the universe in yojanas.

The Pur\={a}\d{n}as provide material which is
believed to be close to the knowledge
of the Vedic times.
Here we specifically consider 
V\={a}yu Pur\={a}\d{n}a (VaP), 
Vi\d{s}\d{n}u Pur\={a}\d{n}a (ViP),
and Matsya Pur\={a}\d{n}a (MP).$^6$
VaP and ViP are generally believed to be amongst the earliest
Pur\={a}\d{n}as and at least 1,500 years old.

The Pur\={a}\d{n}as instruct through myth and 
this mythmaking can be seen in their approach to
astronomy.
For example, they speak of seven underground
worlds
below the orbital plane of the planets
and of seven ``continents'' encircling the Earth.
One has to take care to separate this
imagery, that parallels the conception
of the seven centres of the human's
psycho-somatic body, from the underlying cosmology 
of the physical universe
of the Pur\={a}\d{n}as,
that is their primary concern in their {\it jyoti\d{s}a}
chapters.
The idea of seven regions of
the universe is present in the \d{R}gveda 1.22.16-21 where
the Sun's stride is described as {\it saptadh\={a}man},
or taking place in seven regions.
The geography of concentric continents is
a representation of the plane of the Earth's rotation, with
each new continent as the orbit of the next ``planet''.$^{7}$

The different Pur\={a}\d{n}as 
reproduce the same cosmological material.
There are some minor differences in 
figures that may be a result of wrong 
copying by scribes who did not understand the
material.
In this paper, we mainly follow ViP.

ViP 2.8 describes the Sun to be 9,000 yojanas in
length and to be connected by an axle that is
$15.7 \times 10^6$ yojanas long to the 
M\={a}nasa mountain and another axle
45,500 yojanas long connected to the pole star.
The distance of 15.7 million yojanas between the
Earth and the Sun is much greater than the
distance of 0.46  million yojanas used by
\={A}ryabha\d{t}a in his
Siddh\={a}nta.
(But note that the yojana of the Pur\={a}\d{n}as is
different from the yojana of the Siddh\={a}ntas.)
This greater distance is stated without a corresponding change in
the diameter of the Sun.
The size of the universe is described as 500 million yojanas.

In VaP 50, it is stated that the Sun
covers 3.15 million yojanas in a muh\={u}rta.
This means that the distance covered in a day are 94.5
million yojanas.
MP 124 gives the same figure.
This is in agreement with the view that the Sun is
15.7 million yojanas away from the Earth.
The specific speed given here, translates to 
116.67 yojanas per half-nime\d{s}a.
We have argued that the speed of 2,202 yojanas in
half-nime\d{s}a mentioned by S\={a}ya\d{n}a,$^8$ may
have emerged from the theory that light should
travel at a speed that is able to illuminate the
entire universe in one day.

The size of the universe is described in two different
ways, through the ``island-continents'' and through
heavenly bodies.

The planetary model in the Pur\={a}\d{n}as is
different from that in the Siddh\={a}ntas.
Here the Moon as well as the planets are in orbits
higher than the Sun. 
Originally, this supposition for the Moon may
have represented the fact that it goes higher than the
Sun in its orbit.
Given that the Moon's inclination is $5^\circ$ to
the ecliptic, its declination can be $28.5^\circ$
compared to the Sun's maximum declination of
$\pm 23.5^\circ$.
This ``higher'' position must have been, at
some stage, represented literally by a
higher orbit. To make sense with the
observational reality, it became necessary for
the Moon is taken to be twice as large as
the Sun. 
A point of note is 
that this model effectively assumes that
all the heavenly bodies go around the Sun.

The distances of the planetary orbits beyond the Sun are
as follows:

\vspace{0.3in}
Table 1: From Earth to Pole-star

\begin{tabular}{||l|r||} \hline
Interval I & yojanas\\ \hline
Earth to Sun & 15,700,000\\
Sun to Moon  & 100,000 \\
Moon to Asterisms  & 100,000 \\
Asterisms to Mercury  & 200,000 \\
Mercury to Venus  & 200,000 \\
Venus to Mars  & 200,000 \\
Mars to Jupiter  & 200,000 \\
Jupiter to Saturn  & 200,000 \\
Saturn to Ursa Major  & 100,000 \\
Ursa Major to Pole-star  & 100,000 \\\hline
Sub-total  & 17,100,000 \\ \hline
\end{tabular}
\vspace{0.3in}

Further spheres are postulated 
beyond the pole-star.
These are the Maharloka, the Janaloka, the Tapoloka, and the
Satyaloka. Their distances are as follows:

\newpage
\vspace{0.3in}
Table 2: From Pole-star to Satyaloka

\begin{tabular}{||l|r||} \hline
Interval II & yojanas\\ \hline
Pole-star to Maharloka  & 10,000,000 \\
Maharloka to Janaloka  & 20,000,000 \\
Janaloka to Tapoloka  & 40,000,000 \\
Tapoloka to Satyaloka  & 120,000,000 \\\hline
Grand Total  & 207,100,000 \\ \hline
\end{tabular}

\vspace{0.3in}

Since the last figure is the distance from the Earth, the
total diameter of the universe is 414.2 million yojanas,
not including the dimensions of the various heavenly bodies
and {\it lokas}.
The inclusion of these may be expected to bring this
calculation in line with the figure of roughly 500 million
yojanas.
Or, it is more likely, that the universe is taken to
be egg-like in shape, as suggested by the name of
{\it Brahm\={a}\d{n}\d{d}a}, the world-egg.

Beyond the universe lies the limitless {\it Pradh\={a}na},
that has within it countless other universes.
These other universes were visualized to be independent
world-eggs.

The geography of the Pur\={a}\d{n}as describes a central
continent, Jambu, surrounded by alternating bands of ocean and land.
The seven island-continents 
of Jambu, Plak\d{s}a, \'{S}\={a}lmala, Ku\'{s}a,
Kraunca, \'{S}\={a}ka, and Pu\d{s}kara are encompassed, successively,
by seven oceans; and each ocean and continent is, respectively,
of twice the extent of that which precedes it.

It is important to realize that the continents are
imaginary regions and they should not be
confused with the continents on the Earth.
Only certain part of the innermost continent, Jambu, that deal with
India have
parallels with
real geography.

Although, in earlier work$^9$ we explored the
non-orthodox interpretation that the doubling of
the dimensions applied to the ``oceans'' as well
as ``continents'', we return to the orthodox
view here, where the
increase in dimension by a factor of
two is only across the seven ``continents.''
At the end of the seven island-continents is a region that is twice the preceding
region. Further on, is the Lok\={a}loka mountain, 10,000 yojanas in breadth,
that marks the end of our universe.
The Lok\={a}loka mountain can be compared to the shell of
the world-egg.

Assume that the radius of Jambu is $J_r$ yojana,
then the radius of the universe is:

\begin{equation}
U_r = J_r ( 1 +1 +2+ 2 +2^2 +2^2 +2^3 +2^3 +2^4 +2^4 +2^{5} +2^{5} +2^{6} +2^{6} ) +10,000 
\end{equation}

Or,

\begin{equation}
U_r =  254 J_r + 10,000~ yojanas
\end{equation}

If $U_r$ is roughly 250 million yojanas, 
then $J_r$ should be about 1,000,000 yojanas.

\subsection*{The Orbit of the Sun}

Since the Sun's axle is taken to have dimensions of
$15.7 \times 10^6 $ yojanas, let's see where exactly
it will fit into the island-continent scheme. This is clear when we see that:

\begin{equation}
15.7 \times 10^6 \approx J_r ( 1 + 1 + 2 +  2 + 4 + 4 )
\end{equation}

In other words, the orbit of the Sun will be somewhat
beyond the \'{S}\={a}lmala continent.
Since the total radius of the Universe is
roughly $254 J_r$, it means that beyond the
roughly $14 J_r$ orbit of the Sun, there is present
an approximately 16-fold expansion.

\section*{Y\={a}j\~{n}valkya in B\d{r}had\={a}ra\d{n}yaka U.}

Given the background of the Pur\={a}\d{n}ic 
system, we are ready to examine the statement by
Y\={a}j\~{n}avalkya
in
B\d{r}had\={a}ra\d{n}yaka Upani\d{s}ad (BU) 3.3.2.
He says:
\begin{quote}
{\it dv\={a}tri\d{m}\'{s}atam vai
devarath\={a}hny\={a}nyayam lokasta\d{m} samantam
p\d{r}thiv\={\i} dvist\={a}vatparyeti t\={a}\d{m}
samantam
p\d{r}thiv\={\i} dvist\={a}vatsamudra\d{h}
paryeti}

Thirty-two times the space traversed by the Sun's
chariot in a day makes this plane (loka); around it,
covering twice the area, is the world (p\d{r}thiv\={\i});
around the world, covering twice the area, is the ocean. \hfill (4)
\end{quote}

This describes a system where beyond the Sun's orbit
there is an expansion by a factor of $32$; further beyond that
there is doubling of the area in the dimensions of
the ``Earth'' and a further doubling in the dimensions
of the ocean beyond. 
In other words, there is a total expansion by a factor of
128 beyond the Sun's circuit.

Notice that it is essentially the Pur\={a}\d{n}ic system
in a simplified form. Like the Pur\={a}\d{n}ic
system, the land-mass
and ocean alternate, increasing by a factor of 2 for each
land-mass, although there
is explicit mention in this passage of just one such
region. A second region is the ocean just beyond the Earth
and other similar regions are suggested by the
\d{R}gvedic reference. Y\={a}j\~{n}valkya, collapses several steps by
considering 32 times the space traversed by the Sun,
multiplied by another factor of 4, to be
the Earth's plane.
This system
retains the idea of alternating land and water.

\subsection*{Expansion by Factor of 32}

Assuming that the seven continent-ocean scheme is
meant by the {\it saptadh\={a}man} of the \d{R}gveda,
Y\={a}j\~{n}avalkya, must use it in defining the plane 
that is 32 times the space traversed by the Sun's orbit.
Since beyond this space lie just one continent with
the corresponding ocean (of double the area), the last
continent must be Pu\d{s}kara. Since these two
last regions have a combined size of $128 J_r$,
this plane must be at a distance of
$126 J_r$ from the centre of Jambu.

Let the orbit of the Sun be at a distance
of $S_r$ from the centre of Jambu. Since (4) does
not make it absolutely clear whether the phrase
``space traversed ... in a day'' simply
defines the circuit of the Sun, or whether one should
multiply this circuit by $2 \pi$, we consider
both these possibilities separately.

In the first case, $32 S_r = 126 J_r$, which means
that $S_r \approx 4 J_r$. Or the circuit of the Sun
is just beyond the second continent Plak\d{s}a.
In the second case, $ 2 \pi \times 32 S_r = 126 J_r$.
Or $S_r \approx 0.627 J_r$. In this latter case,
only a part of the Jambu continent is the Earth, and
most of it represents the atmosphere.

If one were to assume that Y\={a}j\~{n}avalkya knew
that $J_r$ was approximately 1,000,000 yojanas, the
radius of the Sun's orbit will then be
4,000,000 yojanas or 626,000 yojanas, respectively.
It is noteworthy, that the Siddh\={a}ntas use
a figure which is not too different from the latter
one.

\section*{Conclusions}

The model of the universe described by
Y\={a}j\~{n}avalkya
in
B\d{r}had\={a}ra\d{n}yaka Upani\d{s}ad (BU) 3.3.2
appears to be the prototype that led to the
full-fledged Pur\={a}\d{n}ic system.
The two systems share essential features of
area doubling and alternating land and water
masses.

There are some difference of details, however.
The Pur\={a}\d{n}ic system seems an expansion
of the Sun's circuit by a factor of 16.
On the other hand, in Y\={a}j\~{n}avalkya's system
the expansion is by a combined ratio of 128.

If it is assumed that 
Y\={a}j\~{n}avalkya
knew the dimensions of the continent and ocean
scheme, the size of the orbit, in one of the
interpretations, comes out to be 
626,000 yojanas, which is only slightly
larger than the number used in the
Siddh\={a}ntas. This suggests that 
Y\={a}j\~{n}avalkya's
scheme was the model from which
both the mature Siddh\={a}ntic and Pur\={a}\d{n}ic
systems emerged.

Even if 
Y\={a}j\~{n}avalkya did not use the
same dimensions as the later Pur\={a}\d{n}ic
astronomy, the structural similarity of the
two systems is striking. 

We find that 
Y\={a}j\~{n}avalkya's scheme moved the Sun
closer to the Earth, compared to the earliest
Vedic scheme where the Sun was right at the
centre of the cosmos. 
It is interesting, nevertheless, that Y\={a}j\~{n}avalkya
projects the Sun's circuit, through his multiplication factor of 32,
to this middle point.
The mature Pur\={a}\d{n}ic system involves an increase
in the size of the Sun's orbit, compared to 
Y\={a}j\~{n}avalkya's scheme.

Y\={a}j\~{n}avalkya's description of
the nature of the cosmos solves a
long-standing puzzle regarding the origin
of the Pur\={a}\d{n}ic system.
We now know that this latter system is
deeply connected to early Vedic ideas.

\section*{Notes}
\begin{enumerate}

\item This work may be found collectively in
S. Kak, ``Birth and early development of Indian astronomy,''
In {\it Astronomy Across Cultures: The History of 
Non-Western Astronomy.} H. Selin (ed.). Kluwer Academic,
Boston, 2000, pp. 303-340;\\ 
S. Kak, {\em The Astronomical Code of the \d{R}gveda}. 
Munshiram Manoharlal, Delhi, 2000.

\item
B.N. Narahari Achar, ``Enigma of the five-year yuga of Vedanga
Jyotisa,'' Indian Journal of History of Science, 33, 1998, pp. 101-109.\\ 
B.N. Narahari Achar, ``On the astronomical basis of the date of
Satapatha Brahmana,'' Indian Journal of History of Science, 35, 2000, pp. 1-19.

\item
S. Kak, ``The astronomy of the \'{S}atapatha Br\={a}hma\d{n}a,''
{\em Indian Journal of History of Science,} 28, 1993, pp. 15-34.\\
S. Kak,
``The astronomical of the age of geometric altars," 
{\em Quarterly Journal of the Royal Astronomical Society,} 36, 1995, 
pp. 385-396.\\
S. Kak,
``Knowledge of the planets in the third millennium BC," 
{\em Quarterly Journal of the Royal Astronomical Society,} 37, 1996, pp. 709-715.

\item 
S. Kak, ``The Sun's orbit in the Br\={a}hma\d{n}as,''
{\em Indian Journal of History of Science,} 33, 1998, pp. 175-191.

\item Note 1.

\item
H.H. Wilson (tr.), {\it The Vishnu Purana}.
Trubner \& Co, London, 1865 (Garland Publishing, New York, 1981);\\
{\it The Matsya Puranam}. The Panini Office, Prayag, 1916 (AMS, New
York, 1974);\\
R.P. Tripathi (tr.),  {\it The V\={a}yu Pur\={a}\d{n}a}. Hindi
Sahitya Sammelan, Prayag, 1987.

\item G. de Santillana, and H. von Dechend, {\it Hamlet's Mill: An Essay
on Myth and the Frame of Time.} Gambit, Boston, 1969.

\item See Note 1.

\item S. Kak, ``Speed of light and Pur\={a}\d{n}ic cosmology.''
In  {\it Computing Science in Ancient India,} T.R.N. Rao and S. Kak (eds.). 
Munshiram Manoharlal, Delhi, 2000.

\end{enumerate}

\end{document}